\documentclass[11pt,a4paper]{article}

\usepackage{graphicx}

\usepackage{cite}

\usepackage{hyperref}

\usepackage{geometry}
\begin{document}
\title{Anisotropic Bouncing Scenario in $F(X)-V(\phi)$ model\\ } 

\date{\today}

\author{Sukanta Panda\footnote{email: sukanta@iiserb.ac.in},
        Manabendra Sharma\footnote{email: manabendra@iiserb.ac.in}}

\maketitle

\centerline{Department of Physics, IISER Bhopal, Bhopal - 462023, India}

\begin{abstract} 
  We investigate the cosmology of a class of model with noncanonical scalar field and matter  in an anisotropic time dependent background. Writing the Einstein Equations in terms of dimensionless dynamical variables appropriately defined for bouncing solutions, we find all the fixed points. From the bouncing conditions and stability of fixed points, solutions describing non singular bounce are  obtained.

\end{abstract}

\section{Introduction}
There are two scenarios exist in literature, namely, inflation and bouncing model which address the shortcomings of the standard model of cosmology.
Though inflation solves most of the problems (horizon, flatness and entropy)
of the standard model of cosmology, the issue with the initial singularity is not resolved under its domain. It is the alternate scenario, nonsingular bouncing model, that eradicates the singularity by constructing a universe which begins with a contracting phase and then bounces back to an expanding phase through a non zero minimum in the scale factor. Nonsingular bouncing models can be categorised into two types, matter bounce model\cite{Fabio} and and Ekpyrotik models\cite{khoury1}. For a review on these scenarios refer to \cite{RHBrev} and \cite{Lehners}. 

A severe problem with bouncing cosmologies is that of instability which develops due to the growth of anisotropic stress during contracting phase, as it grows as sixth inverse power of scale factor. This problem is known as BKL instability\cite{BKL}. 
Originally Ekpyrotic bouncing models were developed to cure this instability\cite{Ekpattr}. This requires a matter field with equation of state parameter $w$ greater than 1. In this case energy density of the matter field dominates over anisotropic stress so that this instability is eliminated.      

  It may be impossible to build a consistent Ekpyrotic bouncing model with one canonical scalar field which is free from BKL instability\cite{khoury1}. The Ekpyrotic scenario given in \cite{khoury1} is not able to produce scale invariant power spectrum with one scalar field. However, the scale invariant spectrum can be achieved by non trivial matching of curvature perturbation across the bounce in an Ekpyrotic scenario\cite{Lyth}. Later, a New Ekpyrotic scenario has been developed, wherein a second scalar field is added which is an isocurvature mode to start with and becomes an adiabatic one during the evolution \cite{khoury2}. This New Ekpyrotic model produces a scale invariant power spectrum. BKL instability develops also in New Ekpyrotic models \cite{xue}. It is shown that the "New Ekpyrotic" scenario can produce nonsingular bounce if the second scalar field is a ghost condensate \cite{arkani}. Unstable growth of curvature  perturbation and anisotropy is still a matter of concern for New Ekpyrotic models \cite{xue}. Whereas matter bounce models can be free from these instabilities as shown in ref. \cite{cai}.

In this paper we consider a noncanonical scalar field with a general function of kinetic term $F(X),$ where $X= - 1/2 \partial_{\mu}\phi \partial^{\mu}\phi.$ These theories are originally motivated to provide a large tensor to scalar perturbation in inflationary settings\cite{Mukhanov:2005bu, Panotopoulos:2007ky, Unnikrishnan:2012zu}. Dark energy with a general kinetic term $F(X)$ is modeled first in ref. \cite{Chiba:1999ka}. For other variants of models of dark energy in this context refer to \cite{Copeland:2006wr}. Other works related to unifying dark matter, dark energy and$/$or inflation for noncanonical scalar field models are studied in \cite{Bertacca:2010ct, Bose:2008ew, Bose:2009kc, DeSantiago:2011qb}. In order to study the phase space in our model, we write the first order equations of motion in terms of dimensionless dynamical variables \cite{Copeland:1997et}. The motivation to use noncanonical scalar field as matter is to construct nonsingular bouncing models. The phase space analysis of a cosmological model with scalar field Lagrangian $F(X)-V(\phi)$ and matter for an FRW background is given in ref.\cite{wands}. The condition for nonsingular bounce is also discussed in ref. \cite{wands}. In order to explore the growth of anisotropy(shear parameter) near bounce and late time isotropization, we do a phase space analysis for the model discussed in \cite{wands} with a Bianchi I metric. This can be easily extended to other nonsingular bouncing models. This analysis may help us to avoid BKL instabilities in bouncing models. It is shown that growth of initial anisotropy and inhomogeneity can be chosen at a level consistent with observations which will be useful for checking the viability of bouncing cosmological models \cite{Garfinkle:2008ei}.

Here, we study the cosmology of an anisotropic universe with a matter Lagrangian of the form $F(X)-V(\phi)$ and an additional matter. In section \ref{EFE} we write the Einstein equations in terms of dynamical variables suitable for studying bouncing scenarios. Fixed points and their stability are analysed in section \ref{FPA}. Conditions for existence of nonsingular bouncing solution, in terms of dynamical variables are derived in section \ref{BouncingScenario}. We summarise our results in section \ref{Conclusion}.

\section{Einstein Equations in Bianchi I background}\label{EFE}
The action for our model is given by
\begin{equation}
S= \int d^4 x \sqrt{-g} [\frac{1}{2} R + F(X) - V(\phi) + L_m] 
\label{}
\end{equation}
where $L_m, ~R, ~g$ and $\phi$ represent the Lagrangian of the matter field, the Ricci scalar, determinant of the metric and the scalar field respectively.

 To see the behaviour of anisotropy of the spacetime in a nonsingular bouncing scenario we choose to work with a homogeneous but anisotropic metric(Bianchi I) with a planar symmetry. The line element of Bianchi I metric with planar symmetry is

\begin{equation}
ds^2=-dt^2+a^2(t)dx^2+b^2(t)(dy^2+dz^2).
\end{equation}



We define average hubble parameter $H$ and Shear $h$ as 

\begin{eqnarray}
H &=& \frac{1}{3}\left ( H_a+2H_b \right )  \nonumber\\
h &=& \frac{H_b-H_a}{\sqrt3} 
\end{eqnarray}

In terms of averaged Hubble parameter and shear h, the Einsteins equation take the following form 
\begin{eqnarray}
\frac{\mathrm{d} H}{\mathrm{d} t} & = &-H^2-\frac{2}{3}h^2-\frac{1}{6}(\rho+3p) \,, \nonumber\\
\frac{\mathrm{d} h}{\mathrm{d} t} & = &-3hH \,, \nonumber\\
H^2 & = &\frac{\rho}{3}+\frac{h^2}{3} \,,
\label{einstein} 
\end{eqnarray}

where $\rho=\rho_{\phi}+\rho_m$ and $p=p_{\phi}+p_m.$

Here the energy density $\rho_{\phi}$ and pressure $p_{\phi}$ of the scalar field is found to be


\begin{eqnarray}
\rho_{\phi}&=&2XF_{X}-F+V \,, \nonumber\\
  p_{\phi}&=&F(X)-V(\phi) \,,
\label{scalarenergydensity}
\end{eqnarray}

and $\rho_m$ and $p_m$ are the energy density and pressure due to $L_m$. 
 
Substituting Eq.(\ref{scalarenergydensity}) in first and third line of Eq.(\ref{einstein}), we get

\begin{equation}
\frac{\mathrm{d} H}{\mathrm{d} t}=-H^2+\frac{2}{3}h^2-\frac{1}{6}(2XF_X-F+V+\rho_m+3(F-V)+3p_m)
\label{rc}
\end{equation}
and 
\begin{equation}
H^2=\frac{2XF_X-F}{3}+\frac{V}{3}+\frac{h^2}{3}+\frac{\rho_m}{3}.
\end{equation}

Here we further define few more variables which are useful for defining dimensionless dynamical variables. They are
\begin{eqnarray}
\rho_k&=&2XF_{X}-F \,, \nonumber\\
w_k&=&\frac{F}{2XF_{X}-F} \,, \nonumber\\
\sigma&=&-\frac{1}{\sqrt3 |\rho_k|}\frac{\mathrm{d} logV}{\mathrm{d} t} \,,
\label{parameters} 
\end{eqnarray}
where $\rho_k$ is the kinetic part of the energy density $\rho_{\phi},$  $w_k$ is the ratio of kinetic part of the pressure $p_{\phi}$ to the $\rho_k$ and $\sigma$ is the auxiliary variable which depends on the variation of potential with time.  

Neglecting the interaction between scalar field and  matter, continuity equation for $\rho_{\phi}$ in terms of dimensionless time variable N $(dN= H dt),$ is
\begin{equation}
\frac{\mathrm{d}}{\mathrm{d} N}(2XF_X-F+V)+6XF_X=0. 
\label{continuity}
\end{equation}

 Now we define a set of  dimensionless dynamical variables which is suitable for nonsingular bounce models. Relevance of these variables are that they remain finite during the entire evolution across bounce. The dynamical variables are

\begin{eqnarray}
\tilde{x}=\frac{\sqrt{3}H }{\sqrt{|\rho_k|}} \,, 
\tilde{y}=\frac{\sqrt{|V|} }{\sqrt{|\rho_k|}}sign(V) \,, 
\tilde{z}=\frac{k}{\sqrt{|\rho_k|}} \,,
\tilde{\Omega}_m=\frac{\rho_m}{|\rho_k|} .
\label{bouncingvariables1}
\end{eqnarray}

Using Eqs.(\ref{bouncingvariables1}), (\ref{einstein}) and (\ref{continuity}) and parameters defined in Eq.(\ref{parameters}), the evolution equations of $\tilde{x},$ $\tilde{y}$ and $\tilde{z}$ are written as,

\begin{eqnarray}
\frac{\mathrm{d} \tilde{x}}{\mathrm{d} \tilde{N}}&=& -\frac{3}{2}\left [ (w_k-w_m)sign(\rho_k)+(1+w_m)(\tilde{x}^2-\tilde{y}\tilde{|y|})+(1-w_m)\tilde{z}^2 \right ]\nonumber\\ &+&\frac{3}{2}\tilde{x}\left [(w_k+1)\tilde{x}-\sigma \tilde{y}|\tilde{y}|sign(\rho_k)  \right ] , \nonumber\\
\frac{\mathrm{d} \tilde{y}}{\mathrm{d} \tilde{N}}&=&\frac{3}{2}\tilde{y}\left [ -\sigma+(w_k+1) \tilde{x}-\sigma \tilde{y}{|\tilde{y}|}sign(\rho_k)\right ] , \nonumber\\
\frac{\mathrm{d} \tilde{z}}{\mathrm{d} \tilde{N}}&=&-3\tilde{z}\tilde{x}+3\tilde{z}\tilde{x}(1+w_k)-3\tilde{z}\tilde{y}|\tilde{y}|sign(\rho_k) , \nonumber\\
\frac{\mathrm{d} \tilde{\Omega}_m}{\mathrm{d} \tilde{N}}&=&-3(1+w_m)\tilde{x}\tilde{\Omega}_m-\tilde{\Omega}_m\left [ 3\sigma \tilde{y}|\tilde{y}|sign(\rho_k)-3\tilde{x}(1+w_k) \right ] \,,  \nonumber\\
\end{eqnarray}

where $d \tilde{N}=\sqrt{\frac{|\rho_k|}{3}} dt$
and the constraint equation relating dynamical variables is
\begin{equation}
\tilde{x}^2-\tilde{y}|\tilde{y}|-\tilde{z}^2-\tilde{\Omega}_m=1\times sign(\rho_k) .
\label{constraint}
\end{equation}

The equation for parameter $\sigma$ becomes \cite{wands}
\begin{equation}
\frac{\mathrm{d} \sigma}{\mathrm{d} \tilde N}= -3\sigma^2\left ( \Gamma-1\right )+\frac{3\sigma\left ( 2\Xi \left ( w_k+1 \right )+w_k-1 \right )}{2\left ( 2\sigma +1 \right )\left ( w_k+1 \right )}\left [ \left ( w_k+1 \right )\tilde x-\sigma \tilde{y}^2 \right ]\, 
\end{equation}
where $\Xi=\frac{X F_{XX}}{F_X}$ and $\Gamma = \frac{V V_{\phi \phi}}{V_{\phi}}.$   
 
For our model we have taken power law form for $F(X)=F_0 X^{\eta},$ where $F_0$ is a constant. For this form of $F(X),$  $w_k=\frac{1}{2 \eta -1}$ and $\Xi=\eta -1.$

Potential $V(\phi)$ is taken as $V(\phi)=V_0e^{-c \phi},$ where $V_0$ and c are constants with positive values. For this choice of $V(\phi),$  $ \Gamma$ becomes unity.   

In the next section, we do a fixed point analysis of dynamical equations for $\tilde x$, $\tilde y$, $\tilde z$ and $\sigma.$ The evolution of $\tilde{\Omega}_ m$ is determined from the constraint Eq.(\ref{constraint}).

\section{Fixed Point Analysis}\label{FPA}

 In this section, we do a fixed point analysis of our system of dynamical equation in order to extract the qualitative information about the nature of solution. Fixed points are calculated by taking the first derivative of the dynamical variables to be zero. The stability of a fixed point is determined from the behaviour of a small perturbation around that  fixed point. 

 We get the set of fixed points $\tilde{x}_c$, $\tilde{y}_c$, $\tilde{z}_c$ and ${\sigma}_c$ by solving the following set of equations simultaneously (where the subscript c denotes fixed points). Now, if we define the slopes of the dynamical variables $\tilde{x}$, $\tilde{y}$, $\tilde{z}$ and ${\sigma}$ as $f(\tilde{x},\tilde{y},\tilde{z},\sigma)$, $g(\tilde{x},\tilde{y},\tilde{z},\sigma)$, $h(\tilde{x}, \tilde{y},\tilde{z},\sigma)$ and $i(\tilde{x},\tilde{y},\tilde{z},\sigma)$. The set of equations we need to solve to obtain the fixed point is

\begin{eqnarray}
f(\tilde{x},\tilde{y},\tilde{z},{\sigma}) &\equiv& \frac{\mathrm{d} \tilde{x}}{\mathrm{d} \tilde{N}}=0 \,,  \nonumber \\
g(\tilde{x},\tilde{y},\tilde{z},{\sigma}) &\equiv& \frac{\mathrm{d} \tilde{y}}{\mathrm{d} \tilde{N}}=0 \,, \nonumber \\
h(\tilde{x},\tilde{y},\tilde{z},{\sigma}) &\equiv& \frac{\mathrm{d} \tilde{z}}{\mathrm{d} \tilde{N}}=0 \,, \nonumber \\
i(\tilde{x},\tilde{y},\tilde{z},{\sigma}) &\equiv& \frac{\mathrm{d} {\sigma}}{\mathrm{d} \tilde{N}}=0 \,, \nonumber \\
\end{eqnarray} 

where,
\begin{eqnarray}
f(\tilde x,\tilde y,\tilde z,\sigma) &\equiv& -\frac{3}{2}[(w_k-w_m)(sign\rho_k)+(1+w_m)(\tilde{x}^2-\tilde{y}|\tilde{y}|)+(1-w_m)\tilde{z}^2]
+\frac{3}{2}\tilde{x}[(w_k+1)\tilde{x}-\sigma  \tilde{y}|\tilde{y}|sign(\rho_k)]\,, \nonumber \\
g(\tilde x,\tilde y,\tilde z,\sigma) &\equiv& \frac{3}{2}\tilde{y}[-\sigma +(w_k+1) \tilde{x} - \sigma \tilde{y} |\tilde{y}| (sign \rho_k)]\,, \nonumber \\
h(\tilde x,\tilde y,\tilde z,\sigma) &\equiv& -3\tilde{z}\tilde{x}+3\tilde{z}\tilde{x}(1+w_k)-3\tilde{z}\tilde{y}|\tilde{y}|sign(\rho_k)\,, \nonumber \\
i(\tilde x,\tilde y,\tilde z,\sigma) &\equiv& \frac{3}{2}\frac{[2 \Xi (w_k+1)+(w_k-1)]}{2(2\sigma+1)(w_k+1)}[(w_k+1)\tilde{x}-\sigma \tilde{y}^2].
\end{eqnarray}

The corresponding fixed point for $\tilde{\Omega}_m$ can be found using the constraint Eq.(\ref{constraint}).

The stability of the fixed points can be examined from the evolution of small pertubations around fixed points. 
Now, if $(\tilde{x}_c, \tilde{y}_c, \tilde{z}_c, {\sigma}_c)$ is a fixed point and $\delta \tilde{x}= \tilde{x}-\tilde{x}_c, \delta \tilde{y}=\tilde{y}-\tilde{y}_c$,
 $\delta \tilde z= \tilde{z}-\tilde{z}_c$  and $\delta \sigma= \sigma-\sigma_c$ be the respective perturbation around it, then the evolution of the perturbation is determined by

\begin{eqnarray}
   \delta \dot{\tilde {x}} &=& \dot{\tilde x}= f(\tilde{x}_c+\delta \tilde {x},\tilde {y}_c+\delta \tilde {y},\tilde {z}_c+\delta \tilde {z},  {\sigma} +\delta {\sigma} ) \,, \nonumber \\
   \delta \dot{\tilde {y}}&=& \dot{\tilde y}= g(\tilde{x}_c+\delta \tilde {x},\tilde {y}_c+\delta \tilde {y},\tilde {z}_c+\delta \tilde {z}, {\sigma}+\delta {\sigma}) \,, \nonumber \\
  \delta \dot{\tilde {z}}&=& \dot{\tilde z}= h(\tilde{x}_c+\delta \tilde {x},\tilde {y}_c+\delta \tilde {y},\tilde {z}_c+\delta \tilde {z}, {\sigma}+\delta {\sigma})\,, \nonumber\\  
  \delta \dot{ {\sigma}}&=& \dot{\sigma}= h(\tilde{x}_c+\delta \tilde {x},\tilde {y}_c+\delta \tilde {y},\tilde {z}_c+\delta \tilde {z}, {\sigma}+\delta {\sigma})\,, \nonumber\\       
\end{eqnarray}

The evolution equations,upto first order, for these pertubations are
\begin{eqnarray}
     {\left( \begin{array}{cccc} \delta \dot {\tilde x} \\ \delta \dot {\tilde y} \\ \delta \dot {\tilde z} \\ \delta \dot \sigma \end{array} \right)    }=  {\bf{A}} {\left( \begin{array}{cccc} \delta {\tilde x} \\ \delta {\tilde y} \\ \delta {\tilde z}  \\ \delta \sigma \end{array} \right)}
\end{eqnarray}

where the matrix is 
\begin{eqnarray}
               {\bf A}=  {\left(
                 \begin{array}{cccc}
                      \frac{\partial f}{\partial \tilde x} & \frac{\partial f}{\partial \tilde y} & \frac{\partial f}{\partial \tilde z} & \frac{\partial f}{\partial \sigma} \\ 
                       \frac{\partial g}{\partial \tilde x} & \frac{\partial g}{\partial \tilde y} & \frac{\partial g}{\partial \tilde z} & \frac{\partial g}{\partial \sigma} \\ 
                        \frac{\partial h}{\partial \tilde x} & \frac{\partial h}{\partial \tilde y} & \frac{\partial h}{\partial \tilde z} & \frac{\partial h}{\partial \sigma}\\
                     \frac{\partial i}{\partial \tilde x} & \frac{\partial i}{\partial \tilde y} & \frac{\partial i}{\partial \tilde z} & \frac{\partial i}{\partial \sigma}
                  \end{array}
                 \right)}
\end{eqnarray}
is the Jacobian matrix and is evaluated at the fixed point $(\tilde{x}_c,\tilde{y}_c,\tilde{z}_c,\sigma_c)$ and hence each entry of $\bf{A}$ is a number. The solution of the system of equations can be found by diagonalizing the matrix $\bf{A}$. A non trivial solution exists only when the determinant $| \bf{A}- \lambda \bf{I}|$ is zero. Thus, solving this equation in $\lambda$ we would get all the eigen values of the system corresponding to each fixed points.

We have two cases: one with  positive kinetic term,  $sign(\rho_k)=+ve$ and other one with negative kinetic term,  $sign(\rho_k)=-ve.$. 

\subsection{Case I, $sign(\rho_k)=+ve$}\label{CaseI}

In this case, we study the fixed points for all possible values of parameters. The fixed point $(0,0,0,0)$ is obtained for $w_k = w_m$ signifying all the dynamical variables $\tilde x$, $\tilde y$, $\tilde z$ and $\sigma$, going to zero at late times. It is a nonhyperbolic fixed point as the eigen value of $\bf{A}$ for this is $(0,0,0,0)$. It's stability cannot be decided from our first order analysis of perturbations. From now onwards eigenvalues would mean eigenvalues of matrix $\bf{A}$ for the rest of the paper.

\begin{figure} 
 \centering
 \includegraphics[width=0.60\textwidth]{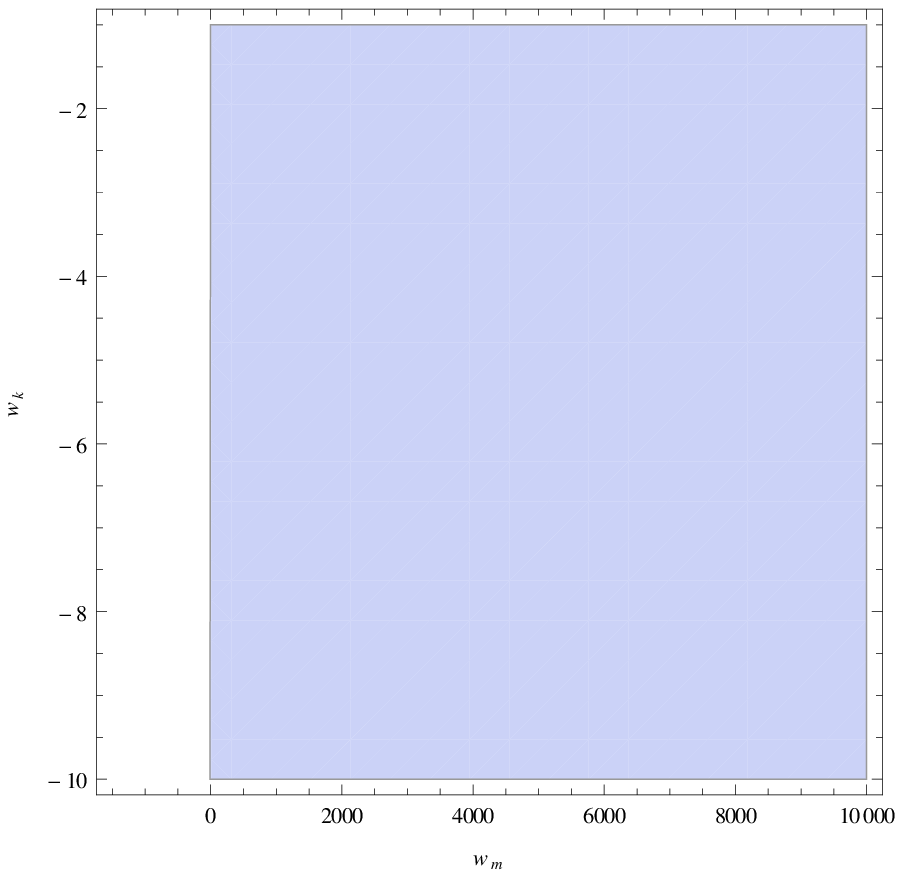}
 \caption{allowed region of parameter space for the fixed point $(1,0,0,0)$}
 \label{ParameterRegionI}
\end{figure}

 The second fixed point  $(1,0,0,0)$ denotes a late time kinetic dominated universe with other dynamical variables $\tilde y$, $\tilde z$ and $\sigma$ becoming zero. In this case, eigenvalue is $(\frac{3(w_k+1)}{2},-3+3(1+w_k),\frac{3}{2}(-1+w_k+\frac{(1-w_k)(1+w_k)}{w_k}),3(1+w_k)-3(1+w_m)).$  This is a stable fixed point for the region of parameter space shown in the Fig. [\ref{ParameterRegionI}]. 

\begin{figure} 
 \centering
 \includegraphics[width=0.60\textwidth]{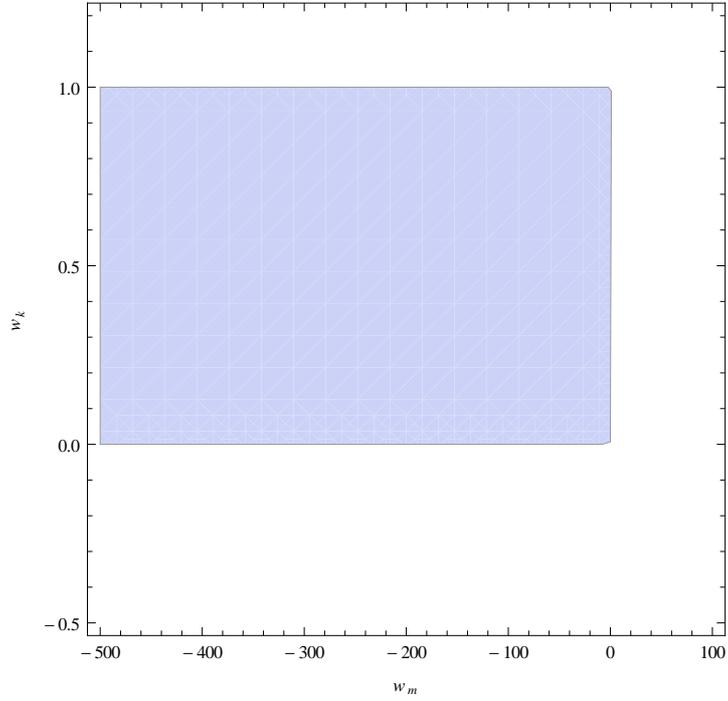}
 \caption{allowed region of parameter space for the fixed point $(-1,0,0,0)$}
 \label{ParameterRegionII}
\end{figure}

The next stable fixed point is  $(-1,0,0,0)$  with eigenvalue $(\frac{3}{2}(-1-w_k),3-3(1+w_k),$\\$ \frac{3(-1-w_k)(-1+w_k+\frac{(1-w_k)(1+w_k)}{w_k})}{2(1+w_k)},\frac{3}{2}(-1-w_k)-\frac{3}{2}(1+w_k)+3(1+w_m))$ shows again a late time kinetic dominated phase but with a negative value of averaged Hubble parameter $H$ signifying a contracting universe. This fixed point 
is found to be stable for the region of parameter space shown in Fig. [\ref{ParameterRegionII}]. The point $(-1,0,0,0)$ may not be important for bouncing point of view, as we need the universe to transit to an expanding phase to be discussed in section \ref{BouncingScenario}. 

The next three fixed points being  $(0,0,-\frac{\sqrt{w_k-w_m}}{\sqrt{-1+w_m}},0)$  for $w_k > w_m$ and $w_m > 1,$ $(0,0,\frac{\sqrt{w_k-w_m}}{\sqrt{-1+w_m}},0)$  for $w_k>w_m$ and $w_m > 1,$  $(0,\frac{\sqrt{w_k-w_m}}{\sqrt{1+w_m}},0,0)$  with eigen values $ (0,0,-3\sqrt{w_k^2-w_kw_m},3\sqrt{w_k^2-w_kw_m})$, $(0,0,-\frac{\sqrt{w_k-w_m}}{\sqrt{-1+w_m}},0)$, $(0,-3\frac{\sqrt{w_k-w_m}}{\sqrt{1+w_m}},-3\sqrt{w_k+w_k^2-w_m-w_kw_m}{\sqrt2},3\frac{w_k+w_k^2-w_kw_m}{\sqrt2}) $ are also nonhyperbolic points. The stability of such fixed points goes beyond the linear stabilty analysis.      
All the fixed points and their stability are noted in table[\ref{tabl:fixpt1}].

\begin{table}
 \centering
 \begin{tabular}{|l|l|l|r|}
  \hline
 Fixed Points $(\tilde{x}_c,\tilde{y}_c,\tilde{z}_c,\sigma_c)$ &              Stability Conditions      \\
  \hline 
   $(0,0,0,0)$ for $w_k=w_m$  &              Can't decide       \\
   $(1,0,0,0)$                &              Stable for $w_k<-1$ and $w_m>0,$ see Fig.[\ref{ParameterRegionI}]     \\
   $(-1,0,0,0)$               &              Stable for $w_k<-1$ and $w_m>0,$ see Fig.[\ref{ParameterRegionII}]     \\
   $(0,0,-\frac{\sqrt{w_k-w_m}}{\sqrt{-1+w_m}},0)$  with $w_k>w_m$ and $w_m>1$  & Can't decide \\
   $(0,0,\frac{\sqrt{w_k-w_m}}{\sqrt{-1+w_m}},0)$  with $w_k>w_m$ and $w_m>1$  & Can't decide \\
   $(0,\frac{\sqrt{w_k-w_m}}{\sqrt{1+w_m}},0,0)$                               & Can't decide \\
  \hline
 \end{tabular}
 \caption{Stablity Analysis of fixed points for $sign(\rho_k)=+1$}
 \label{tabl:fixpt1}
\end{table}

\subsection{Case II, $sign(\rho_k)=-1$}\label{CaseII}

    In this section, we state the results of stability anlysis of our dynamical variables for the negative sign of kinetic energy density. The fixed points are found to be $(0, 0, 0, 0)$, $(0,0,-\frac{\sqrt{-w_k+w_m}}{\sqrt{-1+w_m}},0)$,\\ $(0,0,\frac{\sqrt{-w_k+w_m}}{\sqrt{-1+w_m}},0)$ and $(0,\frac{-w_k + w_m}{1+w_m},0,0)$ with eigen values $(0,0,0,0)$, $(0,0,-3\sqrt{-{w^2}_k + w_k w_m},$\\$3\sqrt{-{w^2}_k + w_k w_m})$, $(0,0,-3 \sqrt{-{w^2}_k+w_k w_m}, 3 \sqrt{-{w^2}_k + w_k w_m})$ and $(0,\frac{\sqrt{-w_k+w_m}}{\sqrt{1+w_m}},0,0)$ respectively. All these fixed points are nonhyperbolic and tabulated in table[\ref{tabl:fixpt2}].

\begin{table}
 \centering
 \begin{tabular}{|l|l|l|r|}
  \hline
 Fixed Points $(\tilde{x}_c,\tilde{y}_c,\tilde{z}_c,\sigma_c)$  & Stability Conditions      \\
  \hline 
   $(0, 0, 0, 0)$ for $w_k=w_m$                      &   Can't decide       \\
   $(0, 0, -\frac{-w_k+w_m}{-1+w_m}, 0)$             &   Can't decide     \\
   $(0, 0, \frac{-w_k+w_m}{-1+w_m}, 0)$              &   Can't decide     \\
   $(0, \frac{-w_k+w_m}{1+w_m}, 0, 0)$               &   Can't decide    \\
  \hline
 \end{tabular}
 \caption{Stablity Analysis of fixed points for $sign(\rho_k)=-1$}
 \label{tabl:fixpt2}
\end{table}

\begin{figure}
 
 $
 \begin{array}{c c}
   \includegraphics[width=0.48\textwidth]{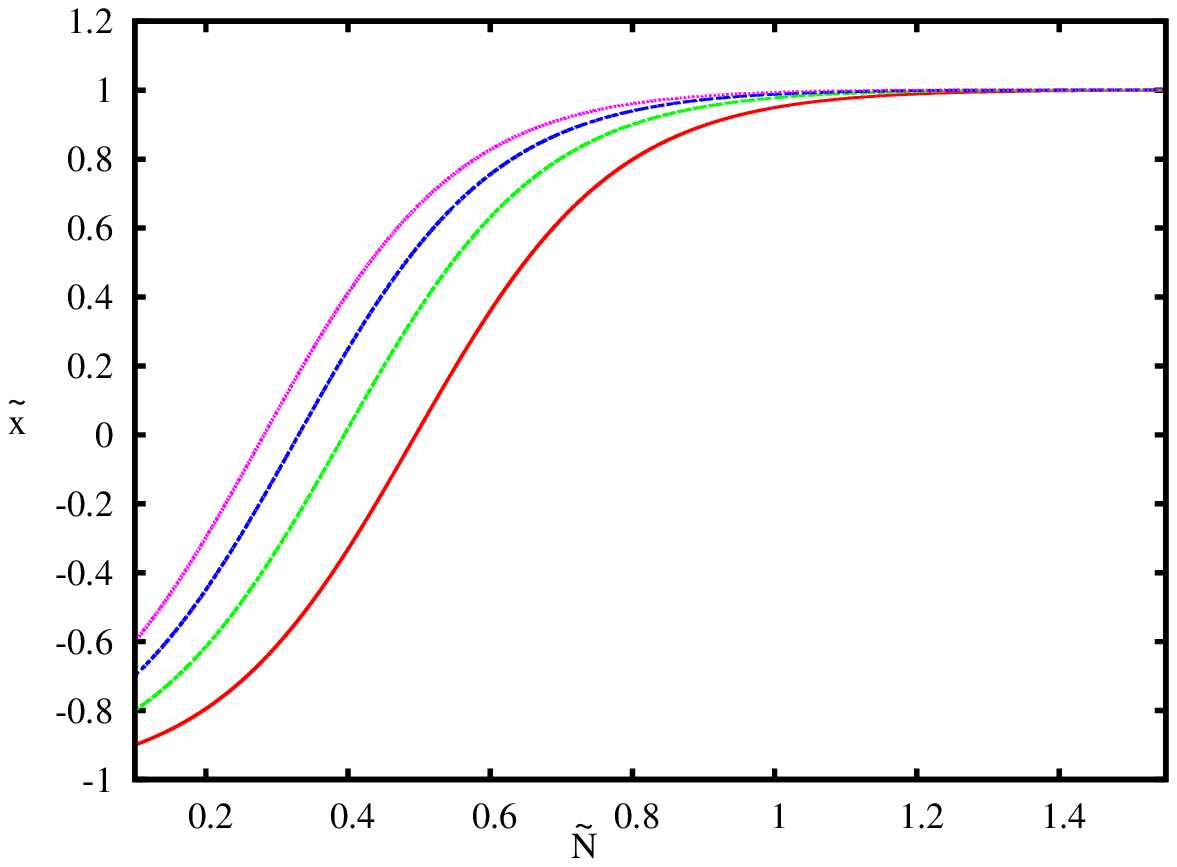} &
   \includegraphics[width=0.48\textwidth]{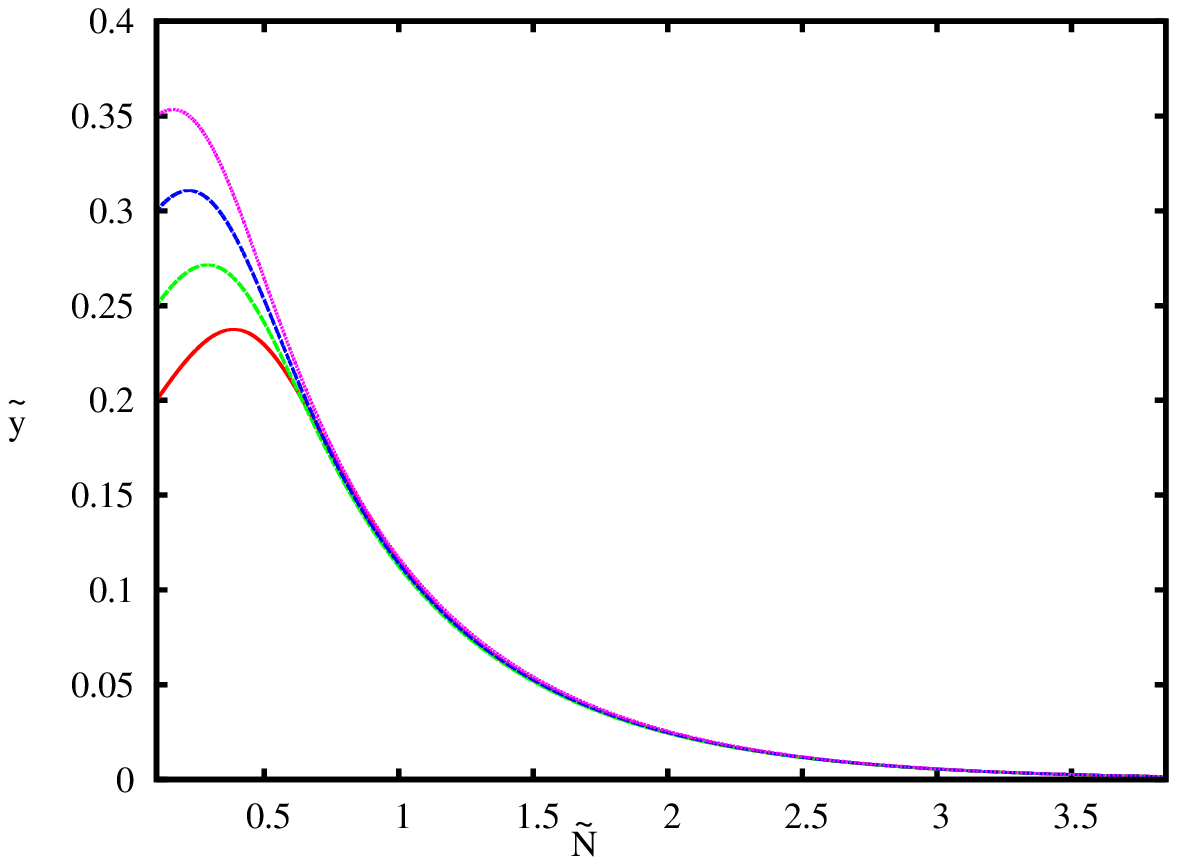} \\
   \includegraphics[width=0.48\textwidth]{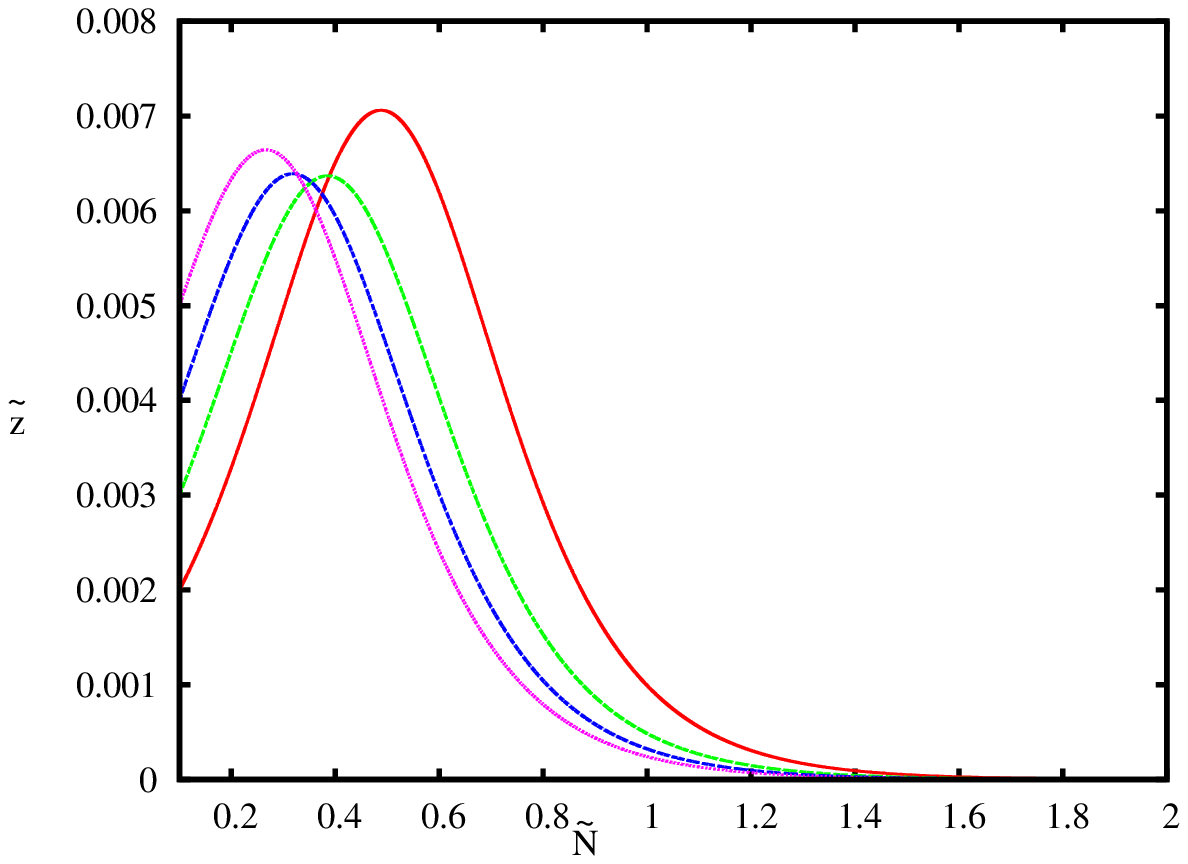} &
   \includegraphics[width=0.48\textwidth]{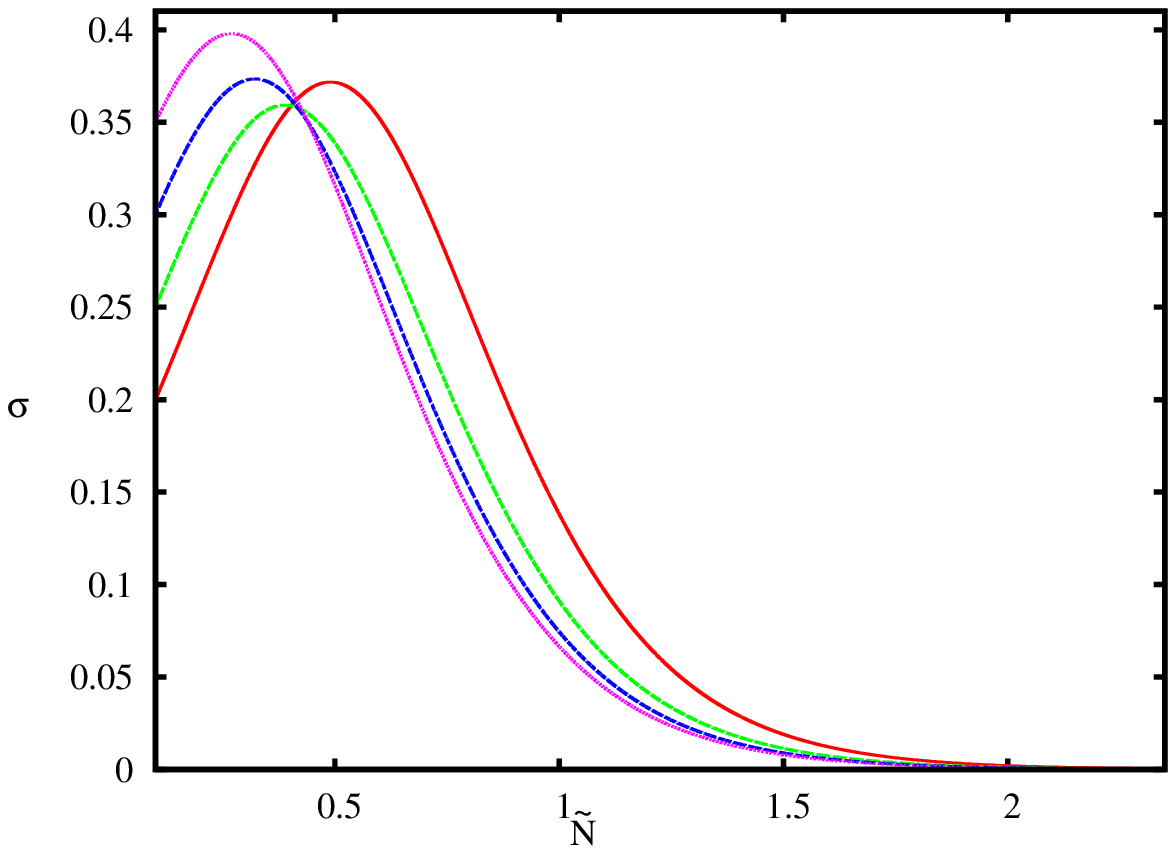}
 \end{array}
 $
  \caption{Evolution of the dynamical variables $\tilde x$ (\emph{top left}), $\tilde y$ (\emph{top right}), $\tilde z$ (\emph{bottom left}) and $\sigma$ (\emph{bottom right}) for the fixed point $(\tilde x_c, \tilde y_c, \tilde z_c, \sigma_c)=(1, 0, 0, 0)$  with the values of parameters $w_k=-2.0$, $w_m=1/3$ and $sign(\rho_k)=+ve$ for different initial conditions}
  \label{fig:FP1000}
\end{figure}

\section{Bouncing Scenario}\label{BouncingScenario}

Now, we obtain the conditions for nonsingular bounce to occur and also show the evolution of dynamical variables numerically. A nonsingular bounce is attained whenever the universe passes from a contracting phase to an expanding phase through a minimum value of the avearge scale factor $A(t)(=(a b^2)^{1/3})$ but not zero. Mathematically, it satisfies
\begin{equation}
(H)_b\equiv \frac{1}{A_b(t)} \left (\frac{\mathrm{d} A(t)}{\mathrm{d}t} \right )_b = 0 ,
\end{equation}
where subscript $b$ denotes value of the variable at the bounce, and 
\begin{equation}
\left ( \frac{\mathrm{d}^2 A(t)}{\mathrm{d} t^2} \right )_b>0
\end{equation}
for minimum to occur. This implies 
\begin{equation}
\left ( \frac{\mathrm{d} H}{\mathrm{d} t} \right )_b=\left ( \frac{\ddot{A}}{A} \right )_b-\left ( \frac{\dot{A}}{A} \right )^2_b>0
\end{equation}

Now, writing the above conditions in terms of dynamical variables for bouncing, we get $\tilde{x}_b=0$  and $\left ( \frac{\mathrm{d} \tilde{x}}{\mathrm{d} \tilde{N}} \right )_b>0$ which translate to the following equations 
\begin{equation}
\left ( \frac{\mathrm{d} \tilde{x}}{\mathrm{d} \tilde{N}} \right )_b
=-\frac{3}{2}\left [ (w_k-w_m)(sign\rho_k)+(1+w_m)(-\tilde{y}|\tilde{y}|)+(1-w_m)z^2 \right ]>0 
\label{bcondition1}
\end{equation}
and
\begin{equation}
\left(\frac{\tilde{y}\tilde{|y|}}{(1-w_m)}-\frac{\tilde{z}^2}{(1+w_m)}\right)_b > 1\times sign(\rho_k)(\frac{w_k-w_m}{1-w_m^2}).
\label{bcondition2}
\end{equation} 

Thus at the bounce, we obtain the constraint equation among dynamical variables as
\begin{equation}
\left ( \tilde{x}^2 -\tilde{y}\tilde{|y|}-\tilde{z}^2-\tilde{\Omega}_m\right )_b = -\tilde{y}\tilde{|y|}-\tilde{z}^2-\tilde{\Omega}_m=1\times sign(\rho_k).
\label{bcondition3}
\end{equation} 

\begin{figure}
 \includegraphics[width=0.50\textwidth]{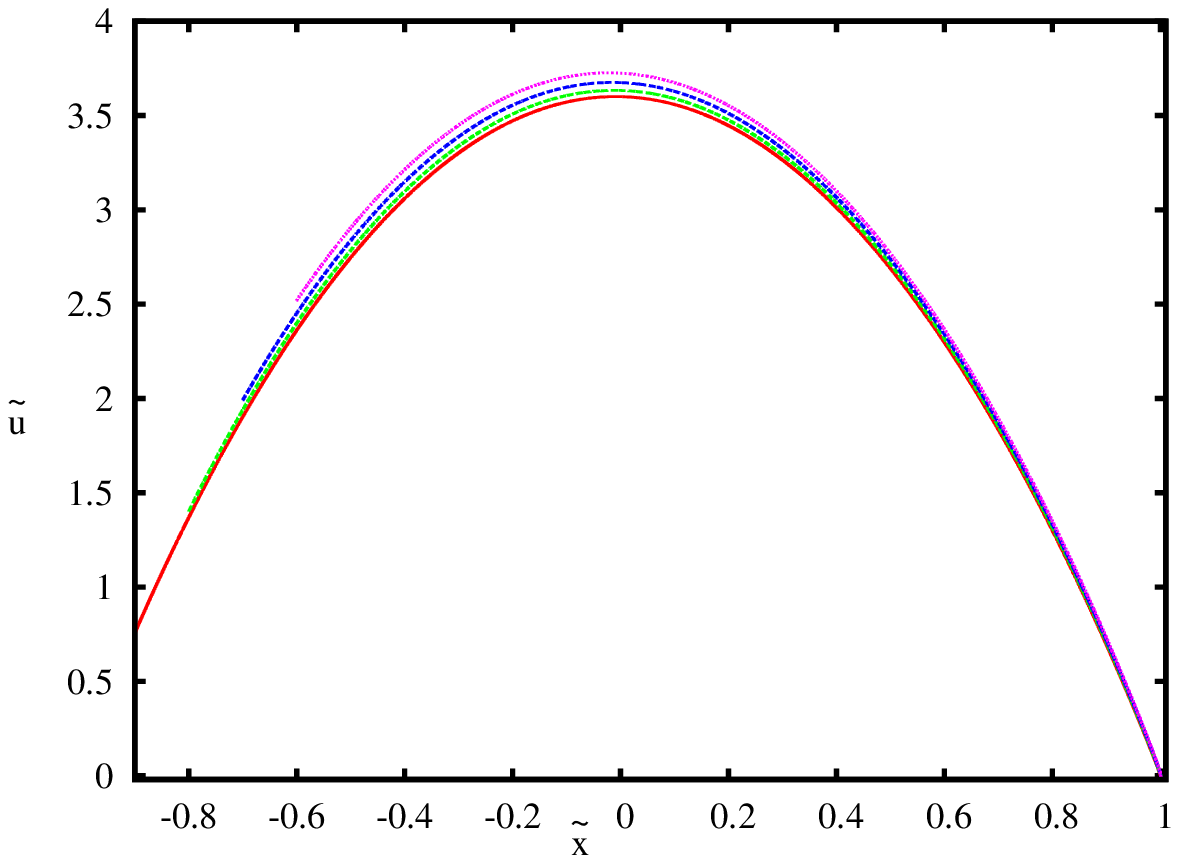} 
 \includegraphics[width=0.50\textwidth]{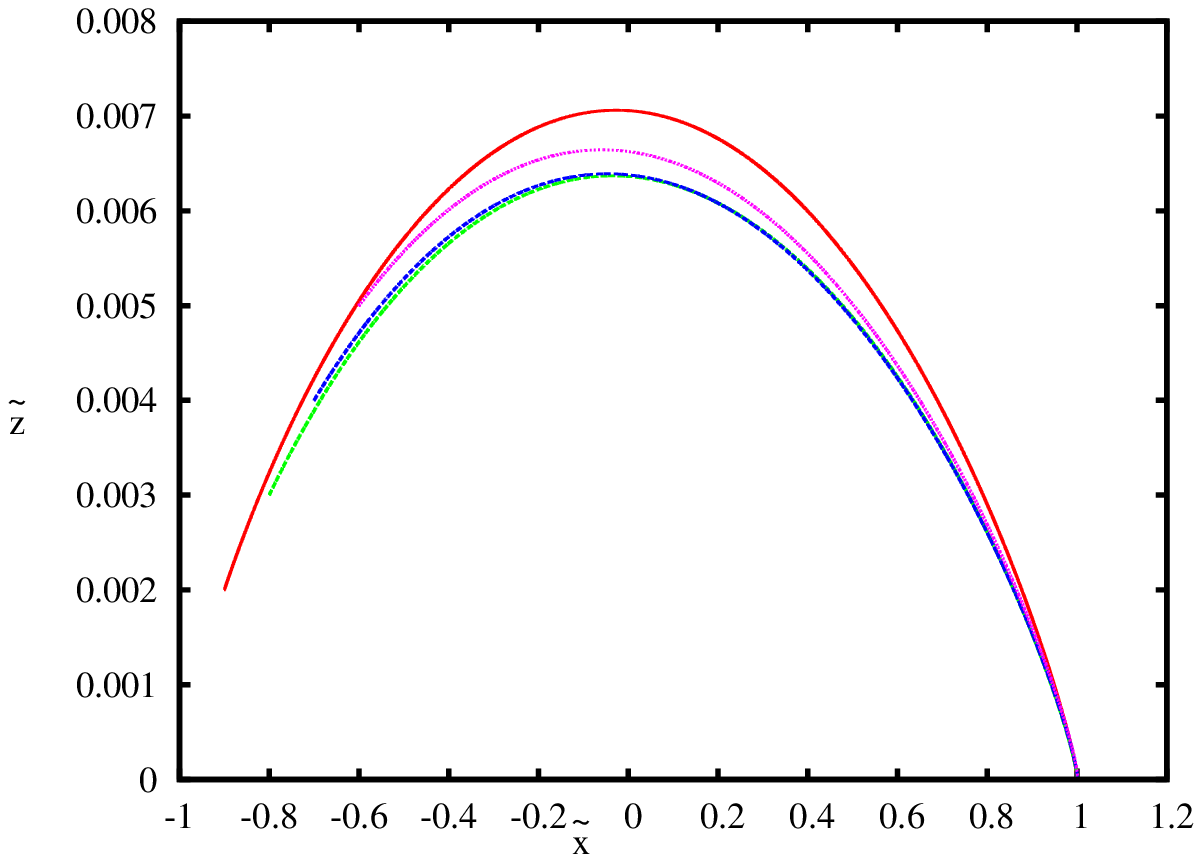}
 \caption{$u\equiv \frac{\mathrm{d} \tilde x}{\mathrm{d} \tilde N}$ vs $\tilde x$ on (\emph{left}) and $\tilde z$ vs $\tilde x$ on (\emph{right}) for $w_k=-2.0$, $w_m=1/3$ and $sign(\rho_k)=+ve$}
 \label{SlopeAndShear}
\end{figure}

          For different negative initial conditions of $\tilde x$ (contracting phase), Fig. [\ref{fig:FP1000}] (\emph{top left}) shows its transition to positive values (expanding phase) crossing zero (bounce). The bouncing is guaranteed by the positivity of the slope of $\tilde x$ as shown in Fig. [\ref{SlopeAndShear}] (\emph{left}). Fig. [\ref{fig:FP1000}] (\emph{top left}) and Fig. [\ref{SlopeAndShear}] (\emph{left}) do indeed represent a stable bouncing scenario. This is obtained by setting the values of equation of state parameters $w_k=-2,$ $(\eta=1/4),$ $w_m = 1/3$ and $sign(\rho_k)=+ve$ and $sign(y)=+ve.$ The evolution of other dynamical variables can be noted from Fig. [\ref{fig:FP1000}], which show their asymptotic evolution to the respective fixed points. 

\begin{figure}
 \centering
 \includegraphics[width=0.84\textwidth]{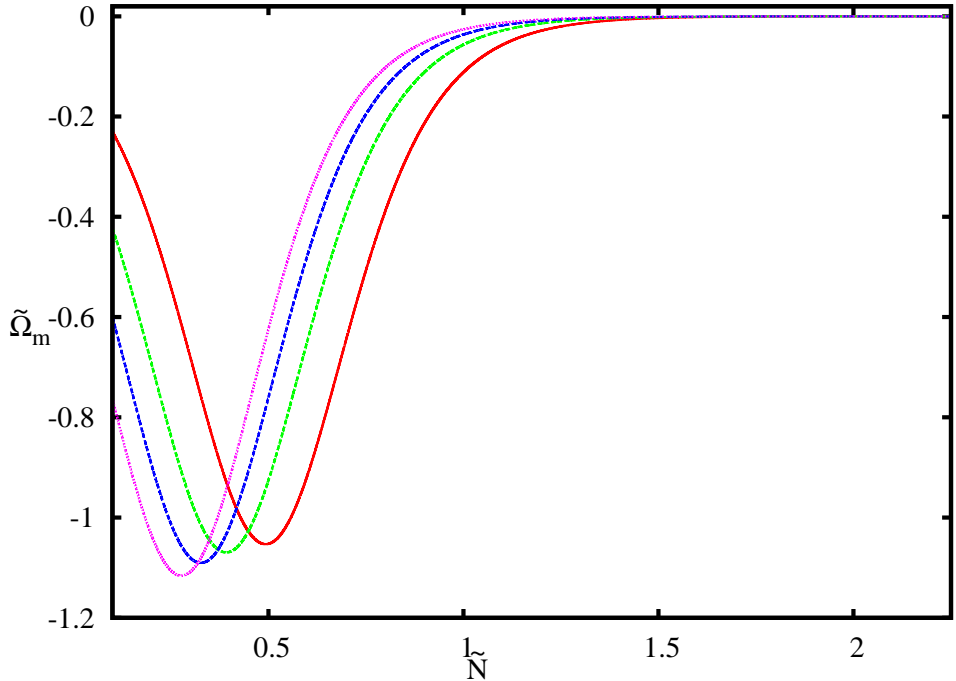}
 \caption{$\tilde{\Omega}_m$ vs $\tilde N$ for $w_k=-2.0$, $w_m=1/3$ and $sign(\rho_k)=+ve$}
 \label{fig:Omega}
\end{figure}

         It can be seen that the fixed point $(\tilde x_c, \tilde y_c, \tilde z_c, \sigma_c)=(1,0,0,0)$ does give rise to a stable bouncing universe provided it satisfies Eqs.(\ref{bcondition2}, \ref{bcondition3}). From this analysis, we conclude that, finally, after the bounce our universe is driven by kinetic energy density at late time. The nonsingular bounce happens only for negative values of $\tilde{\Omega}_m$ with our choice of parameters as shown in Fig.[\ref{fig:Omega}]. The other fixed point $(-1,0,0,0)$, though stable, can not give rise to a bouncing scenario as it ends up with a negative value of averaged Hubble parameter, H, signifying a late time contracting phase. 

          One main point of our work is to show the behaviour of shear parameter, $\tilde z$, in a nonsingular bouncing set up. The shear parameter increases initially as the universe contracts and then decreases to zero value in the expanding phase after crossing the bounce Fig. [\ref{SlopeAndShear}]. Shear parameter remains finite at bounce. Hence, no anisotropic collapse as expected in nonsingular bounce models. Shear may not dominate near the bounce and BKL instability can be avoided which in turn constrain the allowed initial values of dynamical variables.  A late time isotropization of the universe is attained in our model which is useful for building realistic model.

\begin{figure}
 \includegraphics[width=0.50\textwidth]{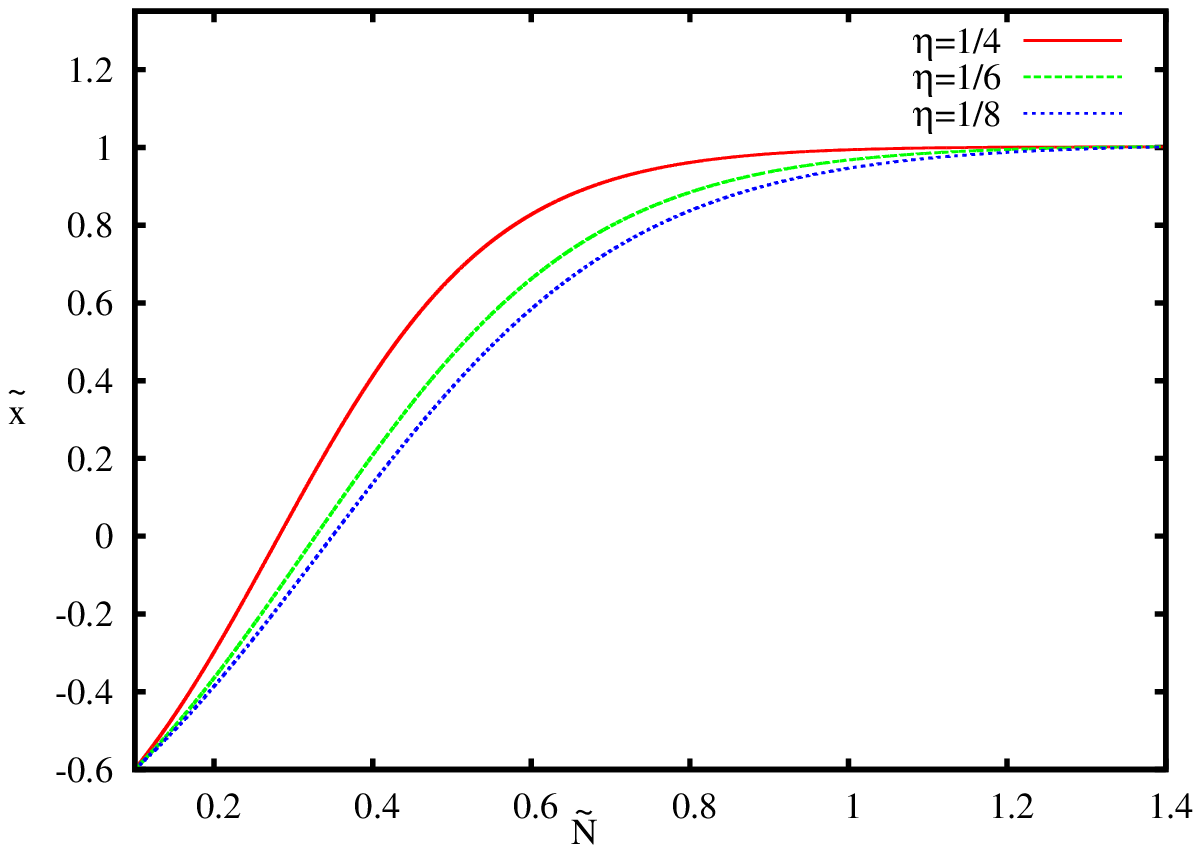} 
 \includegraphics[width=0.50\textwidth]{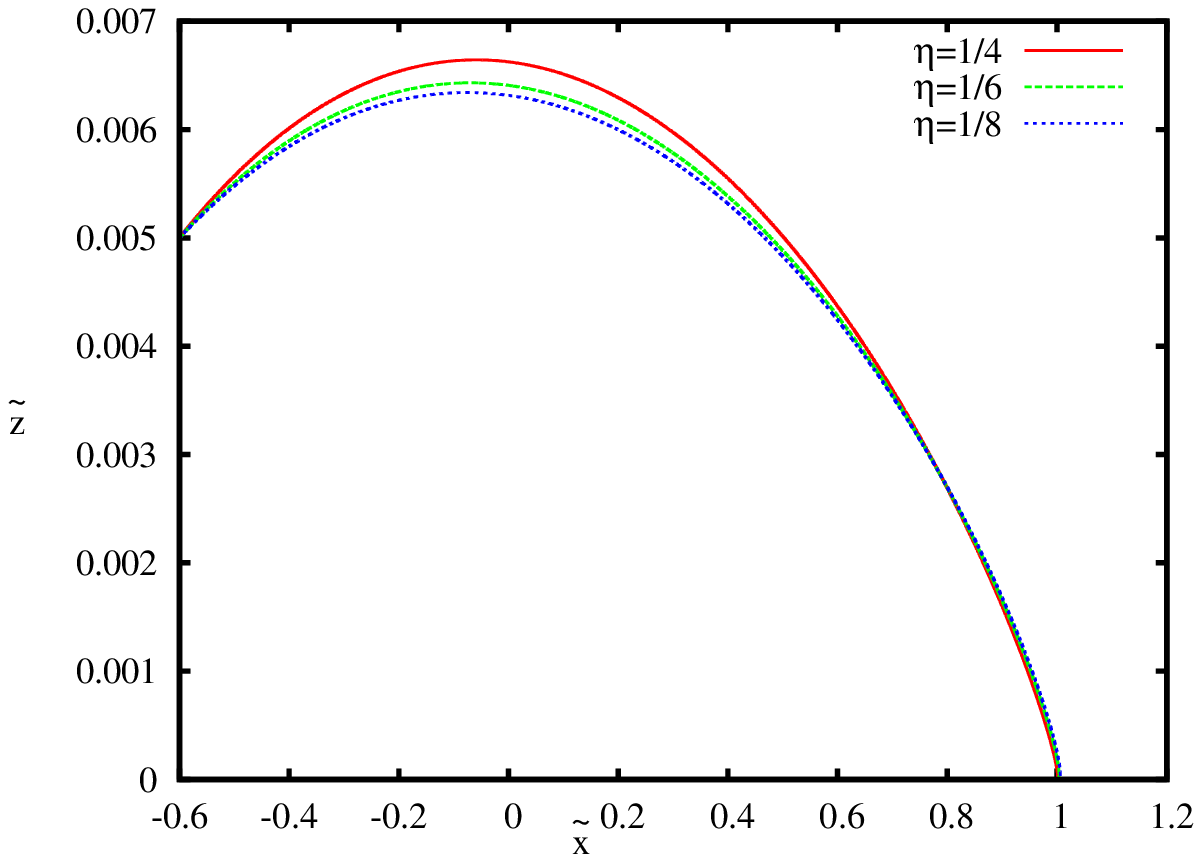}
 \caption{$\tilde x$ vs $\tilde N$ on (\emph{left}) and $\tilde z$ vs $\tilde x$ on (\emph{right}) for $\eta$=$1/4$, $1/6$ and $1/8$ with $w_m=1/3$ and $sign(\rho_k)=+ve$}
 \label{VaryEta}
\end{figure}

         Finally, we show the effect of $\eta$ parameter on the behaviour of bouncing solutions in Fig. [\ref{VaryEta}] (left).  All the plots are generated for the same set of initial conditions and set of parameters $w_m=1/3$, $sign(\rho_k)=+ve$ but with three different values of parameters $\eta=$$1/4$,$1/6$ and $1/8$.  
The values of $\eta$ lies between $0$ and $1/2$ which corresponds to $w_k<-1$ allowed by the stability of fixed points. Thus, higher powers of kinetic term $\eta$ more than or equal to $1/2$ are ruled out by stability criteria. It has been observed that the value of $\eta$ has a direct impact on the occurence of bouncing point. Indeed, the position of bouncing point is delayed as we decrease the value of $\eta$. The right hand side of Fig. [\ref{VaryEta}] shows the variation of shear parameter $\tilde z$ with respect to $\tilde x$ for the three aforesaid values of $\eta$. Though, the universe isotropizes at late time in each of the three cases, it is seen that the value of shear at the bounce is more for higher values of $\eta.$

\section{Conclusion}\label{Conclusion}

A cosmological scenario with a noncanonical scalar field and matter is explored in this work. Using evolution equations for a set of dimensionless dynamical variables, we find all fixed points for both positive and negative kinetic energy density terms. Stability conditions are obtained and allowed region of parameter space is shown. We evolve dynamical variables for different initial conditions with a choice of values of parameters allowed by stability conditions. The necessary and sufficient conditions for a nonsingular bounce to occur in terms of the dynamical variables are derived. Cosmological solutions satisfying both nonsingular bouncing conditions and stability criteria are obtained for the choice of parameters.‏ This is achieved for the negative energy density of matter with equation of state parameter $w_m=1/3.$ In addition to this, the shear parameter is found to be finite at the bounce as expected in a nonsingular bouncing scenario. Late time isotropization occurs in the expanding phase after the bounce which is independent of initial conditions. Finally, the effect of the parameter $\eta$ on the behaviour of bouncing solution is noted. It is seen that the point of occurence of bounce is delayed  as we decrease the value of $\eta$.
 
We restrict our analysis only to positive sign of potential. It is straightforward to extend our analysis for a negative potential by changing the parameter $sign(y)$ to be $-1$. It would be interesting to see the growth of an anisotropic perturbation around Bianchi-I metric in a nonsingular bouncing model and it's signature in cosmological observations.

\end{document}